\documentclass[10pt, conference, compsocconf]{IEEEtran}
\usepackage{multirow}
\ifCLASSOPTIONcompsoc
  \usepackage[nocompress]{cite}
\else
  \usepackage{cite}
\fi

\ifCLASSINFOpdf
  \usepackage[pdftex]{graphicx}
  \usepackage[autostyle]{csquotes}
  \usepackage{caption}
  \usepackage{subcaption}
\else
\fi

\usepackage[cmex10]{amsmath} 
\usepackage{amssymb}
\usepackage{booktabs} % For formal tables

\begin{document}

\title{On the Classification of SSVEP-Based Dry-EEG Signals via Convolutional Neural Networks}

\author{\IEEEauthorblockN{Nik Khadijah Nik Aznan\IEEEauthorrefmark{1}\IEEEauthorrefmark{2},
Stephen Bonner\IEEEauthorrefmark{1}, 
Jason D. Connolly\IEEEauthorrefmark{3}\\
Noura Al Moubayed\IEEEauthorrefmark{1} and Toby P. Breckon\IEEEauthorrefmark{1}\IEEEauthorrefmark{2}}

\IEEEauthorblockA{ Department of $\{$\IEEEauthorrefmark{1}Computer Science, \IEEEauthorrefmark{2}Engineering, 
\IEEEauthorrefmark{3} Psychology$\}$ \\ Durham University, Durham, UK }
}

\maketitle

\begin{abstract}

  Electroencephalography (EEG) is a common signal acquisition approach employed for Brain-Computer Interface (BCI) research. Nevertheless, the majority of EEG acquisition devices rely on the cumbersome application of conductive gel (so-called wet-EEG) to ensure a high quality signal is obtained. However, this process is unpleasant for the experimental participants and thus limits the practical application of BCI. In this work, we explore the use of a commercially available dry-EEG headset to obtain visual cortical ensemble signals. Whilst improving the usability of EEG within the BCI context, dry-EEG suffers from inherently reduced signal quality due to the lack of conduit gel, making the classification of such signals significantly more challenging. 

  In this paper, we propose a novel Convolutional Neural Network (CNN) approach for the classification of raw dry-EEG signals without any data pre-processing. To illustrate the effectiveness of our approach, we utilise the Steady State Visual Evoked Potential (SSVEP) paradigm as our use case. SSVEP can be utilised to allow people with severe physical disabilities such as Complete Locked-In Syndrome or Amyotrophic Lateral Sclerosis to be aided via BCI applications, as it requires only the subject to fixate upon the sensory stimuli of interest. Here we utilise SSVEP flicker frequencies between 10 to 30 Hz, which we record as subject cortical waveforms via the dry-EEG headset. Our proposed end-to-end CNN allows us to automatically and accurately classify SSVEP stimulation directly from the dry-EEG waveforms. Our CNN architecture utilises a common SSVEP Convolutional Unit (SCU), comprising of a 1D convolutional layer, batch normalization and max pooling. Furthermore, we compare several deep learning neural network variants with our primary CNN architecture, in addition to traditional machine learning classification approaches. Experimental evaluation shows our CNN architecture to be significantly better than competing approaches, achieving a classification accuracy of 96$\%$ whilst demonstrating superior cross-subject performance and even being able to generalise well to unseen subjects whose data is entirely absent from the training process.

\end{abstract}

\IEEEpeerreviewmaketitle

%----------------------------------------------------------------------------------------
%	SECTION - Introduction
%----------------------------------------------------------------------------------------
\section{Introduction}

Electroencephalography (EEG) is the most prominent data acquisition approach in BCI, owing to its non-invasive nature, relative ease of use and exquisite temporal resolution \cite{Rao2013, oikonomou2016comparative}. Traditionally, the electrodes used for EEG are placed on the scalp with conductive gel (wet-EEG) in order to lower the impedance between the electrodes and the skin \cite{minguillon2017trends}. The impedance values in EEG signals are a measurement of how good the conductivity is between the electrode and the skin. The lower the value of impedance, the better the electrode and the skin contact thus improving overall EEG signal quality \cite{Lopez-Gordo2014, edlinger2012can}. 

The major drawback of wet-EEG is the required gel application owing to the Ag/AgCl electrodes, consequently resulting in relatively substantial preparation time, scalp discomfort and additional time required to remove the gel after the experimental protocol \cite{edlinger2012can}. Furthermore, the gel will dry over a certain time frame, thus somewhat limiting the experimental data acquisition interval \cite{Lopez-Gordo2014}. Moreover, classical wet-EEG requires some specific experimental conditions like a Faraday cage (a physical shield using conductive material) which reduces the effect of external electromagnetic interference in terms of signal noise \cite{minguillon2017trends}. This limits the application of BCI using wet-EEG to strict experimental operating conditions. By contrast, a dry-EEG headset offers an alternative approach alleviates these limitations in terms of skin preparation, stable connectivity and comfort during experimentation in addition to ready adaptability to different head sizes \cite{mullen2015real, Lin2014}. However, the major drawback is the relatively higher impedance values, as compared to wet-EEG \cite{mullen2013real}, thus making it difficult to reduce the EEG signal noise and unwanted artefacts. This results in a substantially more challenging signal decoding and classification task.

In this study, we are using the commercially available Quick-20 dry EEG headset from Cognionics Inc. (San Diego, USA) with 20 dry-EEG sensors (10-20 sensor layout compliant). The system is employed without the need for skin preparation and it is both portable and wireless \cite{mullen2013real, mullen2015real, Lin2014, lisi2016dry}. This headset comes with individual local active shields that eliminate the need for the rigid experimental condition \cite{Callan2015, mullen2015real}. In our experiments, we collect dry-EEG signals with SSVEP as the neuro-physiological responses. SSVEP has the feature of frequency tagging, which enables the measurement of neural activity in response to a flickering stimuli which the subject is fixated upon, even if the subject is not paying full conscious attention to the stimuli \cite{norcia2015steady}. It is considered to be the most suitable type of stimuli to be used for effective high throughput BCI as SSVEP can provide high Information Transfer Rate (ITR) neural signals with minimal subject training \cite{Kwak2017}.

In this study, we investigate the use of a deep neural network, specifically a CNN, to perform the classification of SSVEP frequencies in dry-EEG data. CNN are a subset of neural networks, which learn to differentiate between classes in data by extracting unique features across multiple layers of convolutional transformation \cite{lecun2015deep}. In the convolution layer, the input is convolved via kernels (filters) to obtain feature maps \cite{Goodfellow-et-al-2016}. This process removes the requirement for hand-crafted feature extraction as well as common signal pre-processing steps, as raw data samples can be used as a direct input to the model \cite{Goodfellow-et-al-2016, schirrmeister2017}. This property provides a critical advantage as the potential exists for salient EEG signals or features to be excluded or missed when using traditional pre-processing based approaches \cite{lawhern2016eegnet}.

We evaluate the performance of our proposed CNN architecture at classifying dry-EEG SSVEP signals across a four class stimuli problem collected from a single subject and highlight the vastly superior performance when compared to baseline classifiers including the Support Vector Machine (SVM), Linear Discriminant Analysis (LDA), Minimum Distance to Mean (MDM) and a Recurrent Neural Network (RNN). Furthermore, we explore the use of the same CNN architecture to examine both multiple subject, exploring both within subject and across subject performance \cite{lawhern2016eegnet}. Finally, to test the ability of the CNN to generalise across \emph{unseen} subjects, we explore the performance when testing upon a subject for which no sample data is present within the training dataset.

In summary, the major contributions of this study are:
\begin{itemize}

\item An end-to-end deep learning CNN architecture to perform the classification of raw dry-EEG SSVEP data without the need for manual pre-processing or feature extraction (the first study to do so with the accuracy achieved: 96$\%$).
\item A demonstrable model that achieves generalisation across subjects during training in contrast to earlier EEG BCI work in the field (accuracy: 78$\%$).
\item An approach with the ability to generalise to entirely unseen subjects with no additional training, raising the potential for subject-independent BCI applications.
\end{itemize}

%----------------------------------------------------------------------------------------
%	SECTION - Related Works
%----------------------------------------------------------------------------------------
\section{Related Work}
% Dry EEG, Cognionics Dry EEG, 
% SSVEP, Frequency SSVEP,
% BCI Deep Learning, - SSVEP using DL 
% Classification per subject  

% Compared to dry-EEG \cite{Lopez-Gordo2014}, wet-EEG requires a long preparation time in order to localise the location of each electrode,  application of the conductive gel to ensure electrode contact, and post session gel removal. The experiment must be carried out in a certain limit of time as the gel will dry out and increase the impedance values of the electrodes. The bulky size and wired connectivity requirements of the wet-EEG also limits the availability of the device to be used in different environments \cite{Liao2014}

In \cite{Lin2014}, a 32-channel dry-EEG was used on subjects in which they fixated on 11 and 12 Hz SSVEP stimuli during walking trials. The performance and the quality of cortical signals between the wet-EEG and dry-EEG during locomotion were compared. From their experiments, wet-EEG performed better as compared to the dry-EEG by 4$\%$ to 10$\%$ in accuracy for standing and walking at different speeds, respectively. 

The study of foot motor imagery has been carried out in \cite{lisi2016dry} to trigger a lower limb exoskeleton while using the same 20-channel dry-EEG headset we use here. The aim of the paper is to have the quick setup system for asynchronous motor imagery BCI as offered by using the dry-EEG headset.

Deep learning approaches have been used in many different BCI applications, akin to motor imagery \cite{schirrmeister2017} as well as the classification of SSVEP signals. In \cite{Kwak2017}, the authors control an exoskeleton via a visual stimulus generator that had five different frequency LEDs to control five different behaviours for static and ambulatory experiments. They used eight wet-EEG electrodes to measure the SSVEP signals with Canonical Correlation Analysis (CCA), Multivariate Synchronization Index (MSI) and CCA with k-Nearest Neighbours (CCA-KNN) used to compare the classification result with three proposed Neural Network (NN) methods: CNN-1 (3 layer network), CNN-2 (4 layer network) and a fully-connected NN. The data from the stimuli is pre-processed for all approaches with the CNN-1 method providing the best accuracy results across both EEG data genres.
 
A five class SSVEP signal problem is classified using both traditional machine learning approaches and deep learning \cite{Thomas2017}. The authors analyse the dataset from the Physionet \cite{goldberger2000physiobank} which used the traditional wet-EEG with five flickering stimuli frequencies. These authors proposed CNN and RNN with Long-short Term Memory (LSTM) for the deep learning methods against traditional classifiers like k-Nearest Neighbour (k-NN), Multi-layer Perceptron (MLP), decision trees and SVM. Within all the classifiers, CNN outperformed other approaches with a mean accuracy of 69.03$\%$ and within the traditional classifiers, SVM provided the best overall accuracy. 

The authors in \cite{lawhern2016eegnet} introduce EEGNet, a CNN model for wet-EEG data across paradigms. The paper includes four datasets for four different paradigms (P300 Event-Related Potential, Error-Related Negativity, Movement-Related Cortical Potential, and Sensorimotor Rhythm). All the datasets come from different sources with different data sizes. These authors pre-process the data before training the datasets using different approaches including both shallow CNN and deep CNN for within subject classification and across subject classification and for all four paradigms. Inconclusively, the results demonstrate that different paradigms perform differently for every approach.  

In contrast to these earlier works, we explicitly consider an end-to-end approach, without the need for EEG signal pre-processing, to tackle single subject, multiple subject and unseen subject SSVEP-based dry-EEG signal classification challenges.

%----------------------------------------------------------------------------------------
%	SECTION - Methodology
%----------------------------------------------------------------------------------------
\section{Methodology}

In this section, we explore the creation of a machine learning model, specifically a deep CNN, in order to perform accurate classification of dry-EEG data. We include several baseline studies in order to compare the performance of the classification accuracy. We also detail the methodology adopted for the experimental SSVEP data collection. 

\subsection{Experimental Setup}
\label{sec:exerimentalsetup}

\begin{figure}[!h]
\centering
\includegraphics[width=0.9\linewidth]{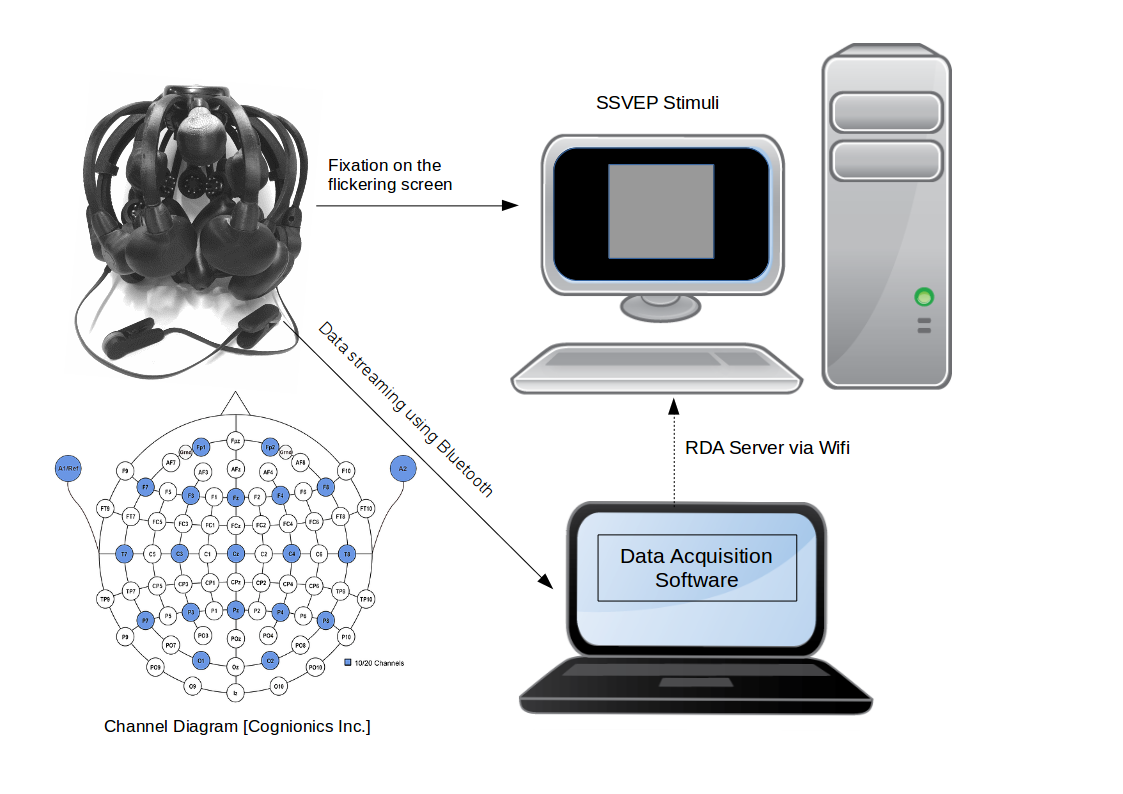}
\caption{Experiment setup and the location of the electrodes of dry-EEG (highlighted in blue)}
\vskip -10pt
\label{fig_sim}
\end{figure}

In this work, we utilise SSVEP as the neuro-physiological response, measured via dry-EEG. The subjects sit in front of a 60Hz refresh rate LCD monitor whilst wearing the dry-EEG headset. We record the data from a range of SSVEP stimuli frequencies; 10, 12, 15 and, 30 Hz \cite{norcia2015steady} using PsychoPy for SSVEP stimuli presentation \cite{peirce2007psychopy}. The stimuli corresponding to the different flicker frequencies were presented on the primary computer. In order to assist with real time processing further along the analysis pipeline, the cortical signals were streamed via the data acquisition software to a secondary computer and sent back to the primary computer. The communication between the different hardware components is shown in Figure~\ref{fig_sim}. 

The dry-EEG headset provides 19 channels and A2, reference and ground as shown in Figure~\ref{fig_sim} (highlighted in blue). The 20-channel (Cognionics Inc.) sensor montage \cite{gargiulo2010new} has been coregistered with the MNI Colin27 brain (Montreal Neurological Institute Colin 27 atlas). Average sensor locations were obtained by averaging 3-D digitized (ELPOS, Zebris Medical GmbH) electrode locations from ten individuals. Electrode labels are assigned based on the nearest neighbour mapping to the standard 10/5 montage. Nas, LPA, and RPA denote nasion and left/right preauricular fiducials \cite{mullen2015real}. 

During the experiments, we collect data over the parietal and occipital cortex (P7, P3, Pz, P4, P8, O1 and O2) \cite{Lin2014}, frontal centre (Fz) and A2 reference at 500 Hz sampling rate across four subjects. The data for subject one (S01) consists of 100 trials of each of the 4 SSVEP classes investigated. For the additional three subjects, we only record 20 trials instead. Each trial flickers the LCD screen for three seconds. The data acquisition software used to monitor and record the signals provides real-time measurement of the impedances for the entire duration of the experiment, thus ensuring good quality EEG signals are recorded.

\begin{figure}[!h]
  \centering
  \begin{subfigure}[b]{0.24\textwidth}
    \includegraphics[width=\linewidth]{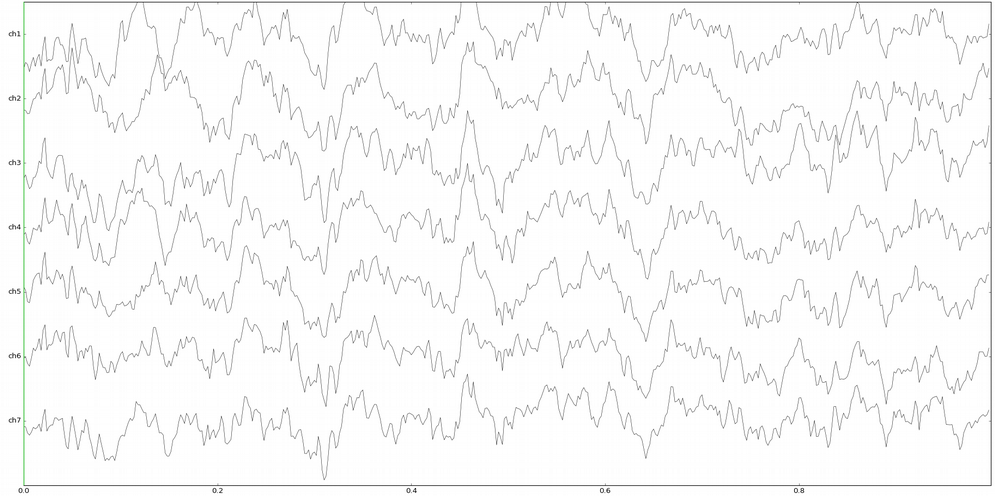}
      \caption{10Hz Signal}
      \label{fig:10hz}
  \end{subfigure}
  \hfill
  \begin{subfigure}[b]{0.24\textwidth}
    \includegraphics[width=\linewidth]{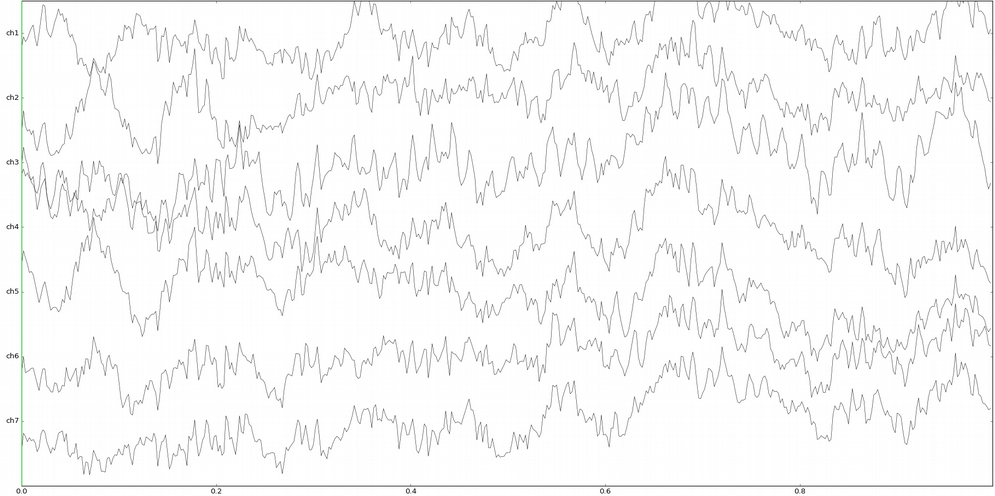}
      \caption{12Hz Signal}
      \label{fig:12hz}
  \end{subfigure}
  
  \begin{subfigure}[b]{0.24\textwidth}
    \includegraphics[width=\linewidth]{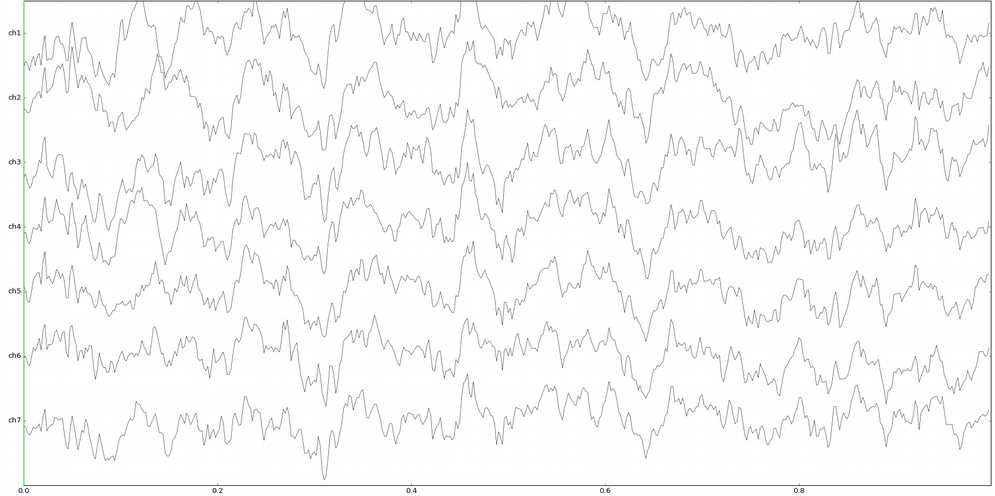}
      \caption{15Hz Signal}
      \label{fig:15hz}
  \end{subfigure}
  \hfill
  \begin{subfigure}[b]{0.24\textwidth}
    \includegraphics[width=\linewidth]{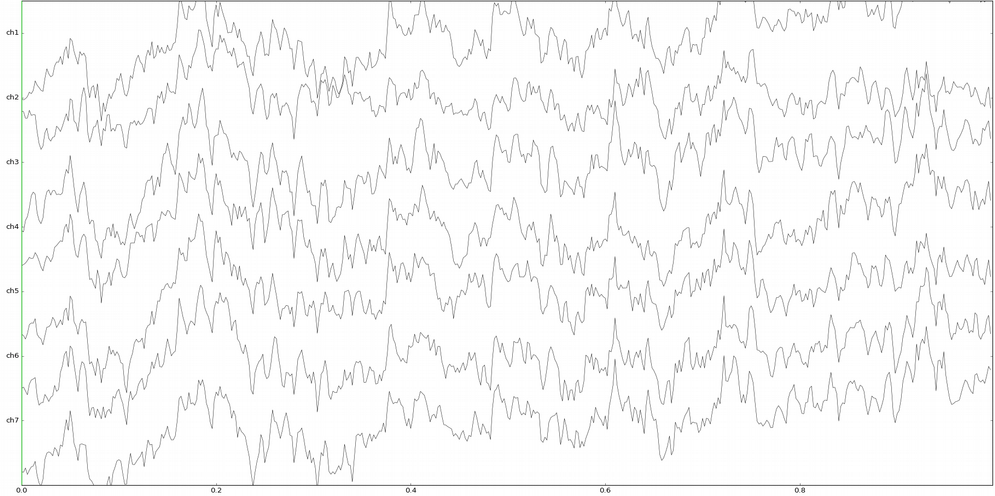}
      \caption{30Hz Signal}
      \label{fig:30hz}
  \end{subfigure}
  \hfill
  \vspace{-0.3cm}
  \caption{Illustrative raw signal data as captured from dry-EEG}\label{fig:frequencyplots}
  \vskip -10pt
\end{figure}

Nevertheless, the primary challenge associated with the classification of dry-EEG signals is the higher noise ratio as compared to the traditional wet-EEG system, owing to the relatively higher impedance values. This noise can be seen in Figure~\ref{fig:frequencyplots} which shows the seven distinct dry-EEG data channels across the four SSVEP frequencies we are investigating. 

\subsection{Convolutional Neural Network Model Design}

Signal processing is one of the primary components in the field of BCI and it acts as the translation between the raw EEG cortical signals to a specific desirable decision or application \cite{minguillon2017trends}. Traditionally, this requires the use of manual pre-processing and feature extraction stages to transform the data into a format suitable for down-stream prediction tasks. By contrast in this work, we explore the use of a deep convolutional neural network to perform this translation process in an end-to-end fashion\footnote{Implemented using the Pytorch library (http://pytorch.org/).}. We explore whether or not a CNN can perform accurate classification of SSVEP target class frequencies on raw dry-EEG data, without the need for manual pre-processing nor feature extraction as found in contemporary work \cite{lecun2015deep}. CNN have demonstrated state-of-the-art results in many image processing tasks, when being used on two dimensional image data \cite{Goodfellow-et-al-2016}. However, there is growing evidence that CNN can be used to process time-series data, when passing a filter over the time dimension, often outperforming recurrent models designed specifically for such temporal data tasks \cite{bai2018empirical}. As EEG data represents time-series data, we make use of a 1D CNN model to classify the dry-EEG data.

\begin{figure}[!h]
  \centering
  \includegraphics[width=0.85\linewidth]{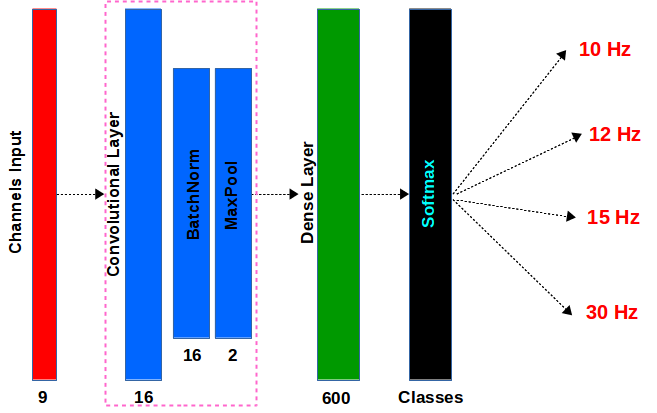}
  \caption{Our proposed 1D CNN architecture including our proposed SSVEP Convolutional Unit (SCU, highlighted in pink)}
  \vskip -10pt
  \label{fig:CNN}
\end{figure}

The structure of the CNN used in this work is displayed in Figure \ref{fig:CNN} in which we have our SSVEP Convolutional Unit (SCU) comprising of a triplicate layer of a 1D convolutional layer, batch normalization and max pooling layer operations. These SCU form the common computational building blocks of the CNN architectures used for dry-EEG signal decoding in this study. Our CNN architecture has a large initial filter to capture the frequencies we are interested in classifying in the dry-EEG data. We also make use of batch normalization to help counterbalance the noisy EEG data. Once the data has been transformed via the convolutional filter, the actual classification of the EEG signal is performed via a softmax function (highlighted in black in Figure \ref{fig:CNN}) in the final layer. The softmax function takes as input the feature vector $x$, generated by the CNN $f_{CNN}(y|x)$ and computes the conditional probability of producing the label $y$ as:

\begin{equation}
  \label{eq:softmax}
 softmax(y|x) = \frac{ \exp{(f_{CNN}(y|x))}}
	% over
 {\sum_{ y^{\prime} \in Y}^{|Y|} \exp{(f_{CNN}( y^{\prime}|x))} },
\end{equation}

where $Y$ is the set of all labels in the dataset.

The loss function the model minimised during training is that of categorical cross-entropy (CCE), which will measure the distance between the output distribution of $\hat y \in f_{CNN}$ and $y \in Y$ as:

\begin{equation}
  \label{eq:crossentropy}
  CCE(y, \hat y) = -\frac{1}{N} \sum_{n=1}^{N} (y_n log (\hat y_n) + (1-y) log (1- \hat y_n)) ,
\end{equation}

where $N$ is the total number of training samples.

 The model is trained using the ADAM gradient descent algorithm \cite{kingma2014adam}, for 100 epochs with a mini-batch size of 32. We also utilise L2 weight decay to help prevent over-fitting by penalising the network for having large weights, meaning that the final objective of our model for optimising is:

 \begin{equation}
  \label{eq:L2}
  Loss = CCE(y, \hat y) + \lambda||f_{CNN}(\mathbf{w})||^2_2 ,
\end{equation}

where $\mathbf{w}$ are the weights of the network and $\lambda$ is a user controllable scaling parameter, set to $10^{-4}$ for this work.

% *CNN Abolition study

\subsection{Baselines}
\label{sec:baselines}

To validate the effectiveness of our proposed approach, we compare with traditional classifiers and other deep learning models. The traditional classifiers used require pre-processing and feature extraction prior to the classification stage. As such, the raw signals will process via the following steps:- downsampled to 250Hz, referencing to the frontal centre sensor signals (Fz), notch filtered at 50Hz to remove line signal noise and bandpass filtered between 9 to 100 Hz. As a result, pre-processing is used to remove the unwanted signals such as power-line noise, and to focus on the signals between the desirable range \cite{minguillon2017trends}. These filtered signals are then utilised as the input for the feature extraction stage. Based on the recent comparative review of \cite{lotte2018review}, we select the Riemannian approach for feature extraction \cite{barachant2013} which utilises covariance matrix and tangent space features which estimate a feature vector in $\mathbb{R}^9$.

Based on the result from \cite{Thomas2017}, SVM is the optimal traditional classifier for EEG data. Therefore we use SVM as one of our baseline classifiers with a Gaussian and linear kernel  \cite{Rao2013}. For further comparison purposes, we also compare with Linear Discriminant Analysis (LDA) and Minimum Distance to Mean (MDM), both frequent choices for EEG analyses \cite{lotte2018review}. To compare with other leading neural network approaches, we also compare with several Recurrent Neural Network (RNN) models \cite{lipton2015critical} including vanilla RNN, Long Short-Term Memory (LSTM) and Gated Recurrent Units (GRU), which have also been assessed for EEG classification in previous study \cite{Thomas2017}.

%----------------------------------------------------------------------------------------
% SECTION - Result and Discussion 
%----------------------------------------------------------------------------------------

\section{Results and Discussion}

In this section, we present detailed experimental evaluation demonstrating the ability of our approach to accurately classify SSVEP signals in dry-EEG data. All the presented results are produced using 10-fold cross validation, with models which were initially optimised using hyperparameters chosen via a grid-search over a validation set. The key hyperparameters which are common across all the networks are L2 weight decay scaling 0.001, dropout level 0.5 and 0.001 for CNN and other deep approaches respectively\footnote{Training time taken for vanilla RNN 600 minutes, GRU 524 minutes, LSTM 619 minutes, CNN 4 minutes on Nvidia GeForce GTX 1060 GPU.}. For the CNN, the hyperparameters utilised are: convolution kernel size 1x10, kernel stride 4, maxpool kernel size 2, and ReLU as the activation function. The dataset and experimental setup are detailed previously in Section \ref{sec:exerimentalsetup}. 

\begin{figure*}[!h]
  \centering
  \begin{subfigure}[b]{0.32\textwidth}
    \includegraphics[width=0.8\linewidth]{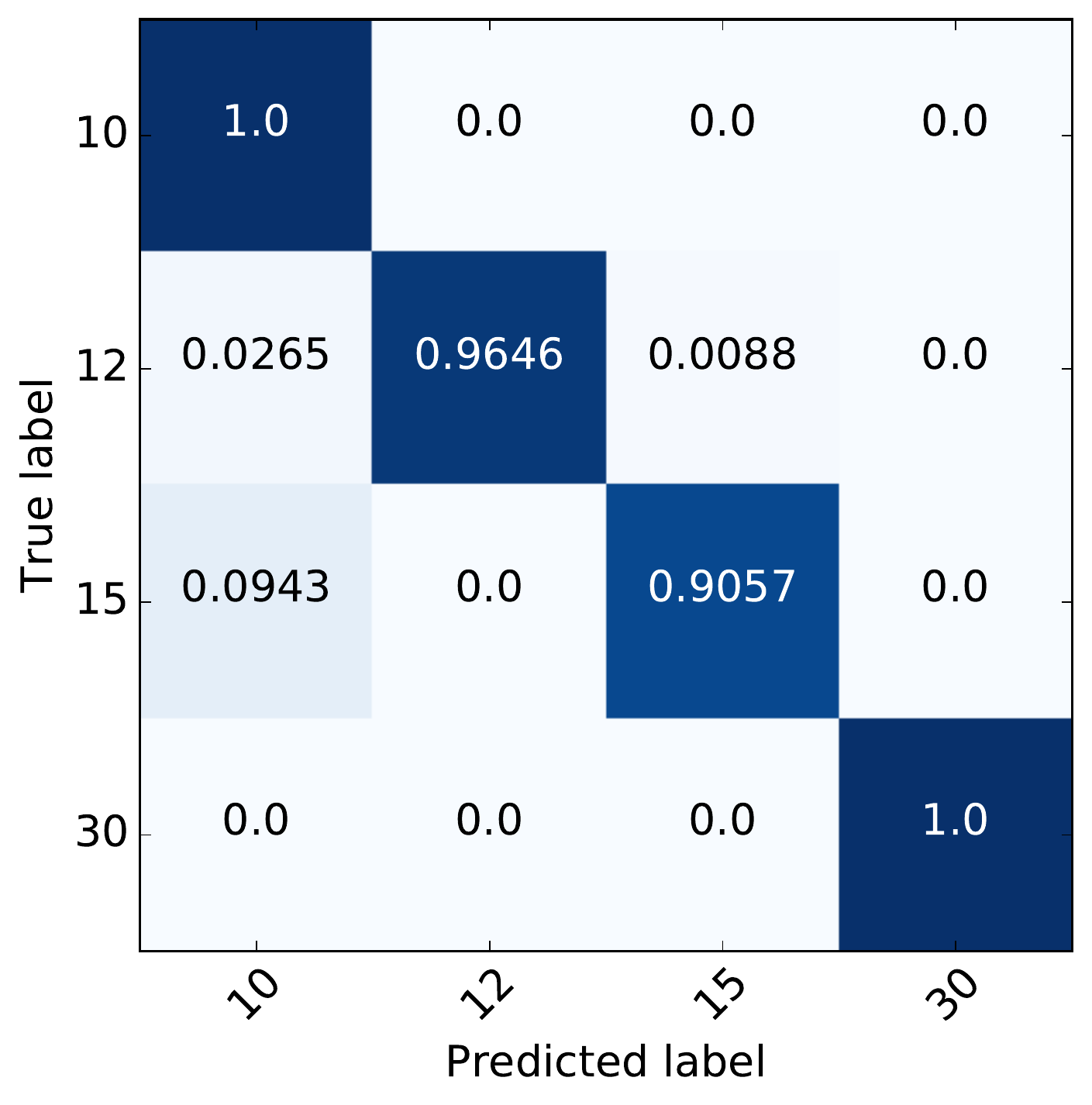}
    \vspace{-0.25cm} 
      \caption{CNN classifying on single subject (S01)}
      \label{CM_S01}
  \end{subfigure}
  \hfill
  \begin{subfigure}[b]{0.32\textwidth}
      \includegraphics[width=0.8\linewidth]{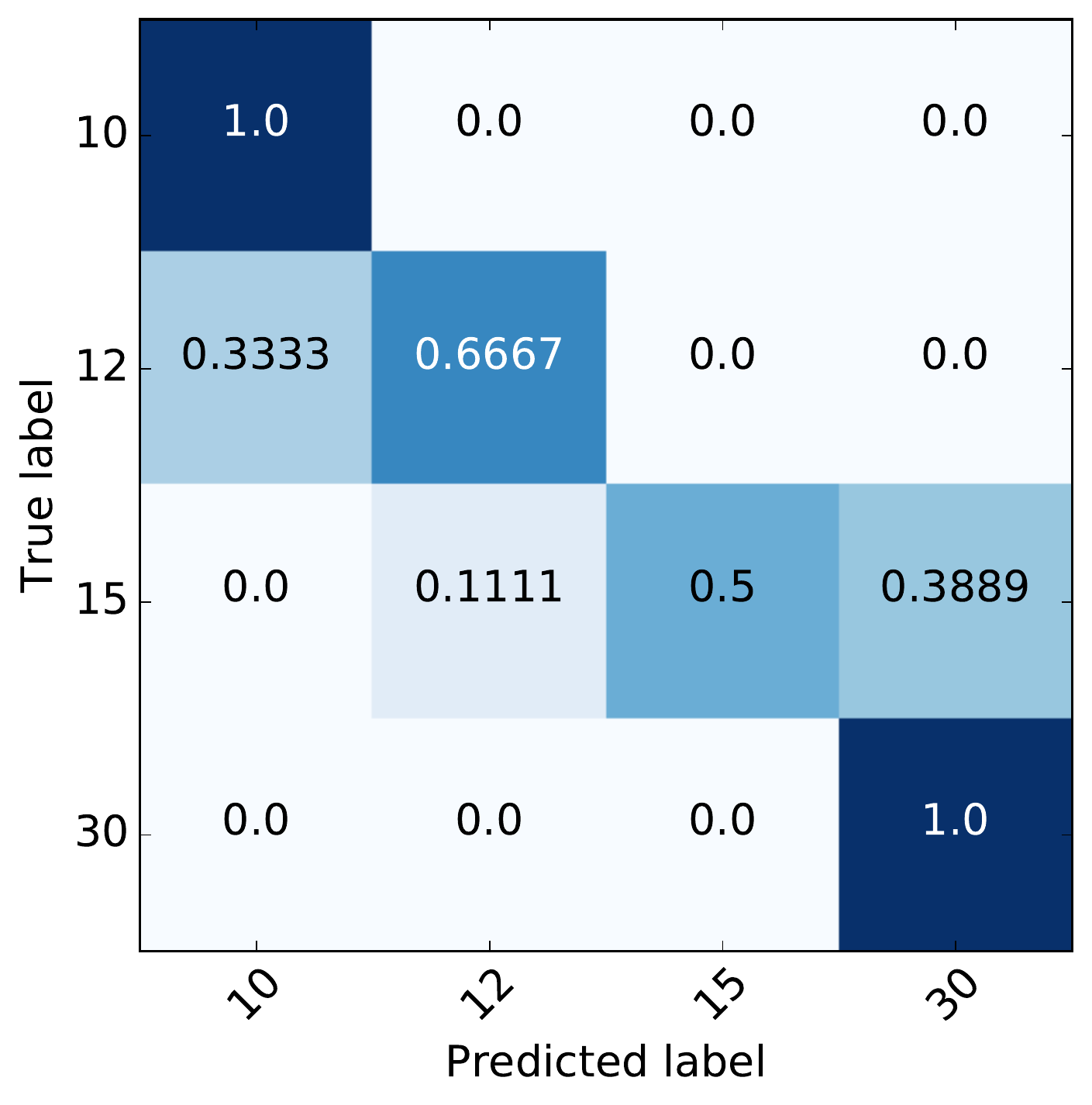}
      \vspace{-0.25cm} 
      \caption{CNN classifying across subjects}
    \label{4subjects_CNN}
  \end{subfigure}
  \hfill
  \begin{subfigure}[b]{0.32\textwidth}
      \includegraphics[width=0.8\linewidth]{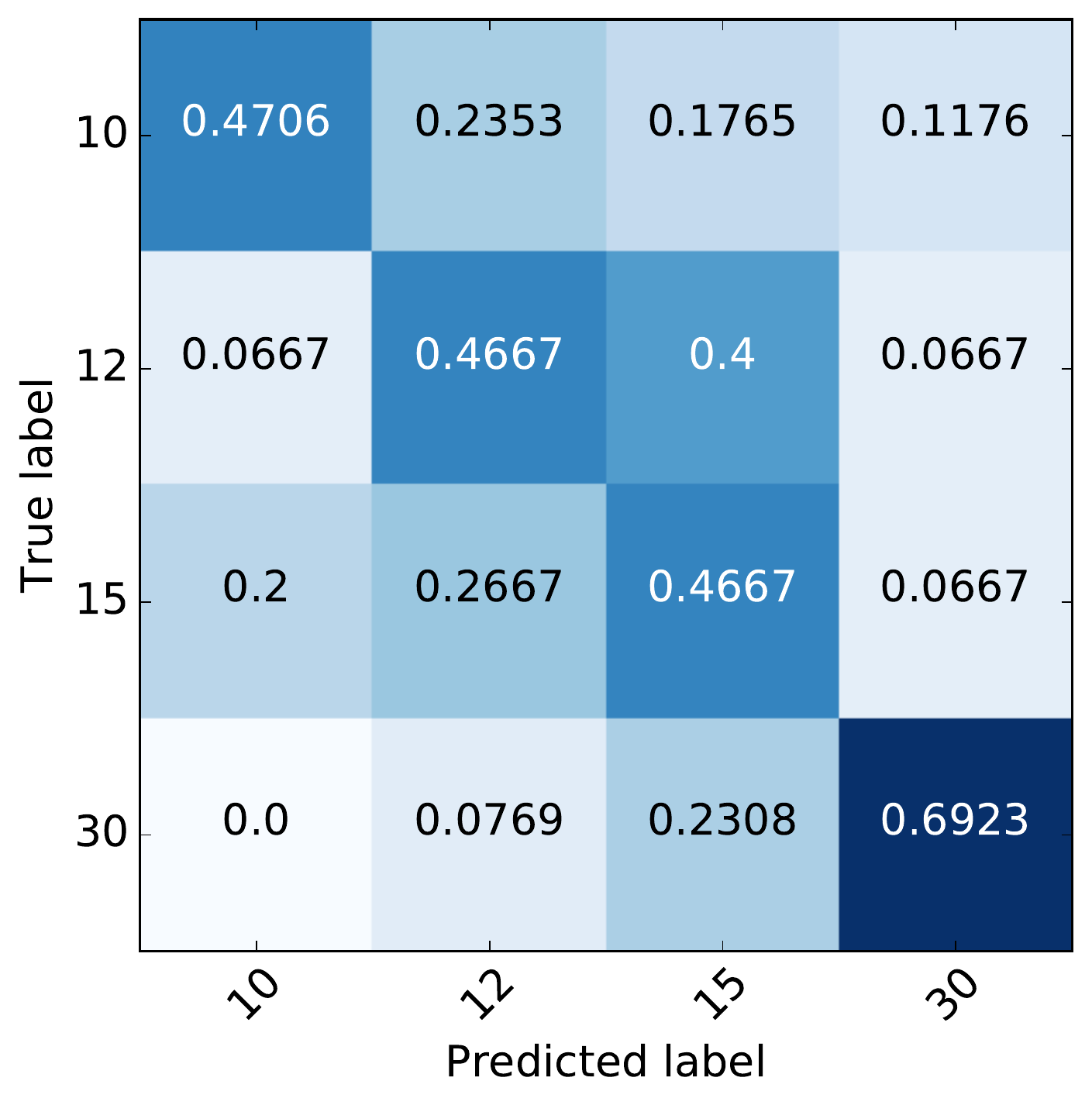}
      \vspace{-0.25cm} 
      \caption{SVM$_{Gaussian}$ classifying across subjects}
    \label{4subjects_SVMG}
  \end{subfigure}
  \vspace{0.05cm}

  \begin{subfigure}[b]{0.32\textwidth}
    \includegraphics[width=0.8\linewidth]{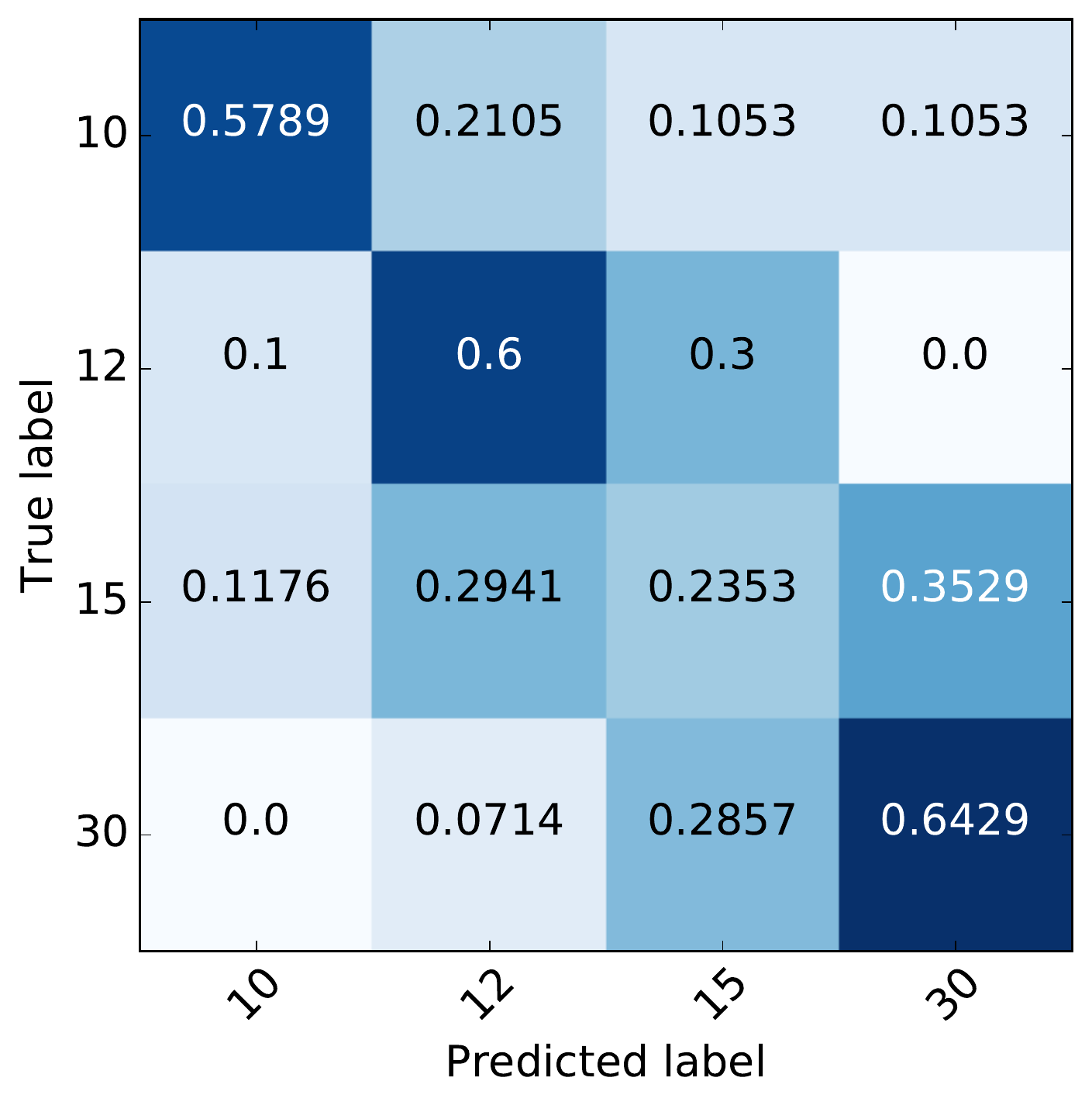}
    \vspace{-0.25cm} 
      \caption{SVM$_{Linear}$ classifying across subjects}
    \label{4subjects_SVML}
  \end{subfigure}
  \begin{subfigure}[b]{0.32\textwidth}
    \includegraphics[width=0.8\linewidth]{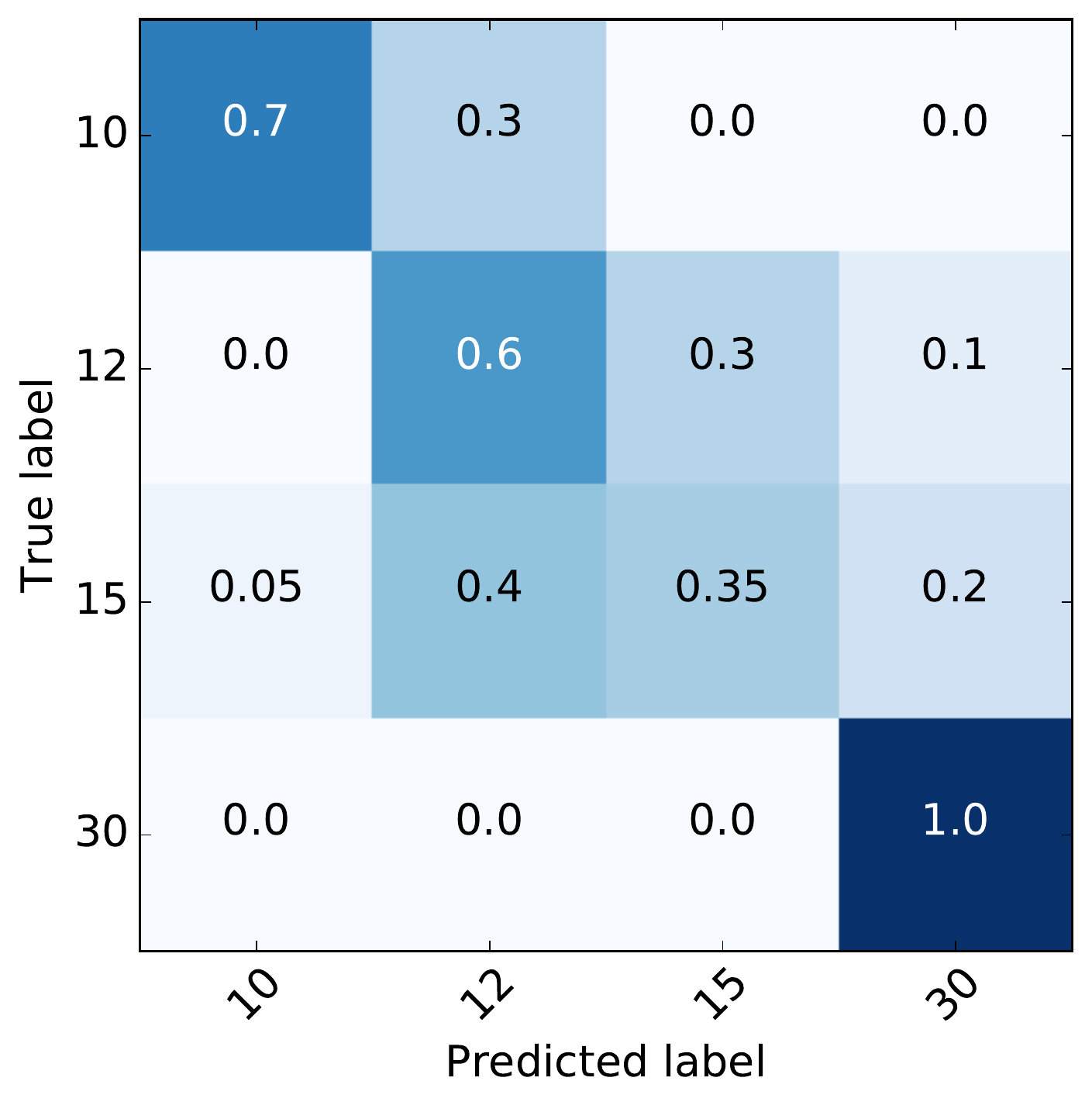}
    \vspace{-0.25cm} 
      \caption{Deep CNN on unseen subject (S04)}
      \label{fig:S04}
  \end{subfigure}
  \caption{Confusion matrices for the various dataset and model configuration tested highlighting the per-class statistical accuracy (Maximal result being \textit{accuracy} = 1.0 in the matrix diagonals). }\label{fig:cm}
  \vskip -10pt
\end{figure*}

\subsection{Single Subject Classification} 
\label{sec:Single_subject}

The results for the classification of the data of a single subject (S01) are presented in Table \ref{one_subject}. The table highlights the accuracy of our proposed CNN approach against all the baselines discussed in Section \ref{sec:baselines}. The results for the traditional approaches, are presented with and without pre-processing. Without pre-processing, we perform feature extraction only before classification with the traditional baseline approaches. Overall, the results show that, even without any pre-processing of the data, our CNN approach demonstrates superior performance over the baselines. The confusion matrix obtained from the classification of S01 using the CNN is presented in Figure \ref{CM_S01}, which shows very strong accuracy across all classes.

\begin{table}[!h]
\centering
\begin{tabular}{c c c c c}
\toprule
\multirow{2}{*}{\textbf{Method}} & \multicolumn{2}{c}{\textbf{Accuracy}} &  &  \\ 
& \emph{Pre-processing} & \emph{Without} &  &  \\ 
\midrule \midrule
CNN & - & \textbf{0.96$\pm$0.02} &  &  \\ 
Vanilla RNN & - & 0.91$\pm$0.05 &  &  \\
LSTM & - & 0.57$\pm$0.30  &  &  \\ 
GRU & - & 0.90$\pm$0.06 &  &  \\ 
SVM$_{Gaussian}$ & 0.92$\pm$0.01 & 0.86$\pm$0.01  &  &  \\
SVM$_{Linear}$ & 0.94$\pm$0.01 & 0.83 $\pm$0.02  &  &  \\ 
MDM & 0.76$\pm$0.01 & 0.80$\pm$0.02 &  &  \\ 
LDA & 0.90$\pm$0.01 & 0.83$\pm$0.01 &  &  \\  
\bottomrule
\end{tabular}
\caption{Mean accuracy with standard deviation over 10-fold cross validation for subject, S01}
\vskip -10pt
\label{one_subject}
\end{table}

\subsection{Multiple Subject Classification} 
 
The second results, presented in Table~\ref{4subjects}, demonstrate the classification performance across three subjects $\{$S01, S02, S03$\}$, where a new classification model is trained for each subject. Due to the known impracticalities of collecting large amounts of data per subject \cite{lotte2018review}, here we reduce the number of SSVEP presentation sessions (trials) per subject for each class to only 20. We also only consider the highest performing classification approaches across traditional and deep model approaches from the previous result of Section \ref{sec:Single_subject} (CNN and SVM; Table \ref{one_subject}). The results highlight, that even with a reduced quantity of data available, the CNN approach still significantly outperforms the SVM across all subjects. This result highlights the applicability of the proposed CNN approach for BCI applications, where data quantity is often relatively limited \cite{lotte2018review}. 

\begin{table}[h!]
\begin{center}
\begin{tabular}{c c c c c c}
\toprule
\textbf{Method} & \textbf{S01} & \textbf{S02} & \textbf{S03} & \textbf{Mean}\\
\midrule \midrule
CNN & 0.91$\pm$0.08 & 0.92$\pm$0.11 & 0.85$\pm$0.10 & \textbf{0.89$\pm$0.03} \\
SVM$_{Gaussian}$ & 0.59$\pm$0.08 & 0.68$\pm$0.08 & 0.67$\pm$0.10 & 0.65$\pm$0.04\\
SVM$_{Linear}$ & 0.76$\pm$0.05 & 0.68$\pm$0.07 & 0.58$\pm$0.10 & 0.67$\pm$0.07 \\
\bottomrule
\end{tabular}
\end{center}
\caption{Mean accuracy with standard deviation over 10-fold cross validation for each of the three subjects, mean results across subjects are also presented}
\vskip -10pt
\label{4subjects}
\end{table}

\subsection{Classification Across Subjects}

To assess the ability of a single CNN model to classify a dataset comprising data from all of the subjects $\{$S01, S02, S03$\}$, we classify all the signals from the three subjects together instead of performing individual classification. Having a single model trained on EEG data from multiple subjects is known to be challenging \cite{dehzangi2018portable}, potentially due to biological differences between subjects and the variability of the EEG recording process. However, the results presented in Table \ref{Generalised} show that the CNN is able to significantly outperform the SVM based approaches when performing classification across subjects. This can be further seen in Figures \ref{4subjects_CNN}, \ref{4subjects_SVMG} and \ref{4subjects_SVML}, showing better performance across classes for the CNN.

\begin{table}[h]
\begin{center}
\begin{tabular}{ c c}
\toprule
\textbf{Method} & \textbf{Accuracy}\\
\midrule \midrule
CNN & \textbf{0.78$\pm$0.10}\\
SVM$_{Gaussian}$ & 0.51$\pm$0.06 \\
SVM$_{Linear}$ & 0.50$\pm$0.06  \\
\bottomrule
\end{tabular}
\end{center}
\caption{Test Accuracy across subjects}
\vskip -20pt
\label{Generalised}
\end{table}

\subsection{Generalisation Capability to an Unseen Subject}

A strongly desirable quality for any model performing the classification of EEG is that of unseen subject generalisation - whereby the model is able to correctly classify data from a subject whose data is absent from \emph{a priori} model training. To test this on our CNN model, we introduce the data of the \emph{unseen} subject S04. We then attempt to classify these data using a model which was trained only on the data of the other three subjects, $\{$S01, S02, S03$\}$. Using the same CNN architecture for this task as depicted in Figure \ref{fig:CNN}, we only achieve an accuracy of 0.59 on S04 without any additional training. We also attempt to classify the new test subject using SVM, however the SVM only displays random classification performance ($\approx$ 0.25 accuracy; i.e 1/4 for 4 classes). 

\begin{figure}[!h]
\centering
\includegraphics[width=0.9\linewidth]{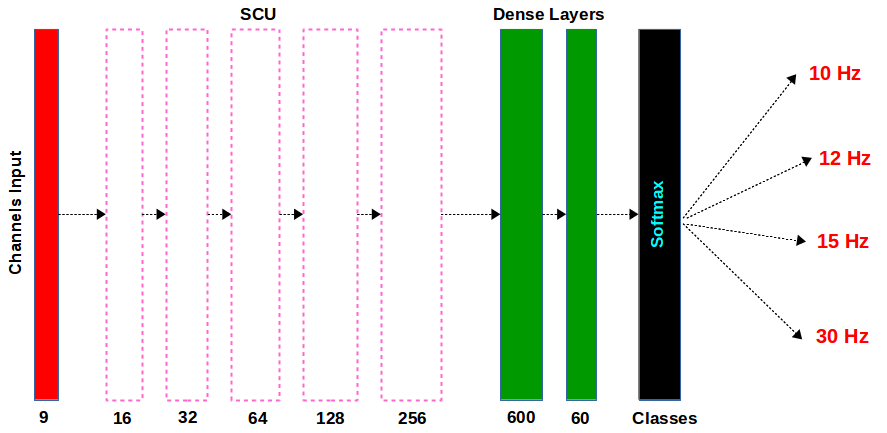}
\caption{Our deeper CNN architecture for \emph{unseen} subjects}
\label{fig:Deep_CNN}
\vskip -20pt
\end{figure}

To overcome this performance issue, we explore a deeper architectural network variant as deeper networks have been shown to learn more complex features in order to determine the correlation between subjects \cite{Goodfellow-et-al-2016}. Figure~\ref{fig:Deep_CNN} illustrates the deeper architecture where empirically, we repeat our SCU blocks (each dashed pink box represents a SCU block) to a maximum number of five. This deeper architecture, introduced to classify subject S04 data, demonstrates a substantially better classification accuracy of 0.69, perhaps suggesting that a deeper model is required to perform the unseen subject generalisation task. The confusion matrix for this result is presented in Figure~\ref{fig:S04}. This figure demonstrates that the CNN has varying performance across the different classes, with the 30Hz signal being the best performing for this extended CNN model. 

%%----------------------------------------------------------------------------------------
%    SECTION - Conclusion
%----------------------------------------------------------------------------------------
\section{Conclusion}

In this paper we introduce deep convolutional neural network architectures constructed around a common computational building block, for the classification of raw dry-EEG SSVEP data - the first such study to do so. We evaluate the performance of our model on SSVEP data recorded from four subjects using the noise-prone dry-EEG methodology. As compared with current state-of-the-art methods, our approach requires no pre-processing to the data, demonstrates higher overall classification accuracy across subjects and generalises significantly better to entirely unseen test subjects. These key results demonstrate that CNN based approaches should become the new benchmark method for SSVEP dry-EEG classification.

Future work would involve larger datasets to further study the classification and generalisation performance across subjects. The combination of the CNN and RNN models may also offer a way to increase overall performance. 

\ifCLASSOPTIONcompsoc
  \section*{Acknowledgments}
\else
  \section*{Acknowledgment}
\fi
The authors would like to thank the Ministry of Higher Education Malaysia and Technical University of Malaysia Malacca (UTeM) as the sponsors of the first author.

\bibliographystyle{IEEEtran}
\bibliography{ref}

\end{document}